\documentclass[conference]{IEEEtran}
\IEEEoverridecommandlockouts
\usepackage{cite}
\usepackage{amsmath,amssymb,amsfonts}
\usepackage{algorithmic}
\usepackage{graphicx}
\usepackage{textcomp}
\usepackage{xcolor}
\usepackage{tcolorbox}
\usepackage{balance}
\usepackage{listings}
\usepackage[T1]{fontenc}
\usepackage{beramono}
\usepackage{tikz}
\usepackage{hyperref}
\usepackage{url}
\usepackage{makecell}
\usepackage{multirow}
\usepackage[flushleft]{threeparttable}
\usepackage{ragged2e}
\makeatletter 
\newcommand{\linebreakand}{%
  \end{@IEEEauthorhalign}
  \hfill\mbox{}\par
  \mbox{}\hfill\begin{@IEEEauthorhalign}
}
\makeatother
\definecolor{blue}{HTML}{000000}
\definecolor{violet}{HTML}{000000}
\definecolor{teal}{HTML}{000000}
\definecolor{red}{HTML}{0000B9}

\def\BibTeX{{\rm B\kern-.05em{\sc i\kern-.025em b}\kern-.08em
    T\kern-.1667em\lower.7ex\hbox{E}\kern-.125emX}}

\begin{document}
\newcounter{todocounter}
\newcommand{\giang}[1]{{\color{red} giang:  #1}}

\title{Demystifying Faulty Code with LLM: Step-by-Step Reasoning for Explainable Fault Localization
}

\author{\IEEEauthorblockN{Ratnadira Widyasari}
\IEEEauthorblockA{\textit{Singapore Management University} \\
ratnadiraw.2020@phdcs.smu.edu.sg}
\and
\IEEEauthorblockN{Jia Wei Ang}
\IEEEauthorblockA{\textit{Singapore Management University} \\
jiawei.ang.2022@scis.smu.edu.sg}
\and
\IEEEauthorblockN{Truong Giang Nguyen}
\IEEEauthorblockA{\textit{Singapore Management University} \\
gtnguyen@smu.edu.sg}
\and
\linebreakand
\IEEEauthorblockN{Neil Sharma}
\IEEEauthorblockA{\textit{Singapore Management University} \\
neil.sharma.2022@scis.smu.edu.sg}
\and
\IEEEauthorblockN{David Lo}
\IEEEauthorblockA{\textit{Singapore Management University} \\
davidlo@smu.edu.sg}
}

\maketitle

\thispagestyle{plain}
\pagestyle{plain}

\begin{abstract}
Fault localization is a critical process that involves identifying specific program elements responsible for program failures. Manually pinpointing these elements, such as classes, methods, or statements, which are associated with a fault is laborious and time-consuming. To overcome this challenge, various fault localization tools have been developed. These tools typically generate a ranked list of suspicious program elements. However, this information alone is insufficient. A prior study emphasized that automated fault localization should offer a rationale. 

In this study, we investigate the step-by-step reasoning for explainable fault localization. We explore the potential of Large Language Models (LLM) in assisting developers in reasoning about code. We proposed FuseFL that utilizes several combinations of information to enhance the LLM results which are spectrum-based fault localization results, test case execution outcomes, and code description (i.e., explanation of what the given code is intended to do). We conducted our investigation using faulty code from \textit{Refactory} dataset. First, we evaluate the performance of the automated fault localization. Our results demonstrate a 32.3\% increase in the number of successfully localized faults at Top-1 compared to the baseline. To evaluate the explanations generated by FuseFL, we create a dataset of human explanations that provide step-by-step reasoning as to why specific lines of code are considered faulty. This dataset consists of 324 faulty code files, along with explanations for 600 faulty lines. Furthermore, we also conducted human studies to evaluate the explanations. We found that for 22 out of the 30 randomly sampled cases, FuseFL generated correct explanations. 
\end{abstract}


\section{Introduction}
\label{sec:intro}
Fault localization, a necessary process in debugging, is aimed at localizing the specific code elements responsible for failures~\cite{wong2016survey}.
 Given the time-consuming nature of manual fault localization, many previous studies proposed approaches for automated fault localization techniques. These proposed techniques to localize the fault encompass a range of methods, including machine-learning-based approaches~\cite{widyasari2022xai4fl, sohn2017fluccs}, static analysis-based approaches~\cite{neelofar2017improving, feyzi2018fpa}, spectrum-based approaches~\cite{wong2013dstar, jones2001visualization,abreu2006evaluation,abreu2009spectrum,naish2011model}, and so on. These techniques typically rank program elements from the most to the least suspicious.

However, it is often not sufficient to only rank program elements based on their suspiciousness (i.e., likelihood to be the root cause of the failures). Providing an explanation for a particular likely faulty location, as highlighted by a previous study~\cite{kochhar2016practitioners}. Their survey shows that 85\% of the respondents agree or strongly agree that fault localization needs to provide a rationale, explaining why a particular code element is flagged as buggy by the tool. To address this need, there are several ways to offer a rationale.  
One of them is through step-by-step reasoning, which is an approach that involves breaking down complex reasoning into a sequence of logical steps or actions, making it easier for one to follow and understand the thought process. 
It has been highlighted by previous studies that step-by-step reasoning is a valuable tool for enhancing rationality and understanding, whether in scientific investigations, ethical considerations, or even debates~\cite{shepard2008step,dennett2013intuition}. 

In recent years, Large Language Models (LLM) have become increasingly popular due to their capability to generate natural and coherent text. There have been several usages of LLM in the software engineering community with promising results, including automated program repair~\cite{xia2023automated, liu2023refining}, code generation~\cite{feng2023investigating}, and vulnerability detection~\cite{fu2023chatgpt}. 
\textcolor{blue}{In recent studies by Wu et al.~\cite{wu2023large} and Kang et al.~\cite{kang2023preliminary}, they utilize LLM and proposed prompts for fault localization.  They show a higher number of successfully localized faults compared to the baselines using the Defects4J dataset.  
However, these studies~\cite{wu2023large, kang2023preliminary} did not include an evaluation of the explanations generated by LLM. Consequently, we lack insight into the quality of these generated explanations. 
Suitable explanations are important to convince practitioners on the trustworthiness of LLM-powered tools, which affects their adoption in practice~\cite{lo2023trustworthy}.
This knowledge gap could be attributed to the absence of fault localization datasets that provide explanations for why specific code elements are considered faulty.  To fill this gap, we aim to use LLM to generate explanations to aid debugging and evaluate to what extent they are good.}

In this study, we investigate the usage of LLM for automated fault localization. We instructed LLM to provide step-by-step reasoning on why the specific line is deemed as faulty. We utilize a combination of information related to the fault, including the spectrum-based fault localization (SBFL) results, test case execution outcomes, and code description. We refer to this approach as FuseFL. 
To assess the quality of the FuseFL-generated explanations, we curated a dataset comprising 324 faulty code files sourced from the \textit{Refactory} dataset~\cite{hu2019re}. While the faulty code files were taken from \textit{Refactory} dataset~\cite{hu2019re}, we manually generated the explanations for the faulty lines. For each faulty line in the code, we write its respective step-by-step rationale, explaining why that particular line was classified as faulty.

We compared FuseFL with 5 widely recognized SBFL techniques~\cite{pearson2016evaluating} (i.e., Tarantula~\cite{jones2001visualization}, Ochiai~\cite{abreu2006evaluation}, OP2~\cite{naish2011model}, Barinel~\cite{abreu2009spectrum}, and DStar~\cite{wong2013dstar}), XAI4FL~\cite{widyasari2022xai4fl}, and LLM-based approach proposed by Wu et al.~\cite{wu2023large}. FuseFL outperforms SBFL (Ochiai) by 68.4\% and XAI4FL by 32.3\% in localizing fault at the Top-1 position. 
Furthermore, FuseFL also achieved a 31.3\% increase in successfully localized faults at Top-1 compared to the LLM-based approach~\cite{wu2023large} (see Section~\ref{sec:discussion_prompt}). 

Comparing the human explanation with the FuseFL-generated explanation, we got a BLEURT~\cite{sellam2020bleurt} (Bilingual Evaluation Understudy with Representations
from Transformers) score of 0.492 for the Top-1 explanation. BLEURT is a metric designed to evaluate natural language generation tasks, which is based
on BERT~\cite{devlin2018bert}. We found that the seemingly low BLEURT score does not necessarily signify an incorrect explanation but, rather, implies a variance between the FuseFL-generated explanation and the human explanation.  
\textcolor{blue}{For example, the FuseFL-generated explanation tends to focus on the suggestion to fix the code while our human explanation primarily focuses on diagnosing the code issue and describing the nature of the error. The FuseFL-generated explanation's emphasis on providing corrective action and its use of different terminology and structure compared to the human explanation may contribute to the observed disparity in BLEURT scores.} 

In addition, we also conduct human studies to evaluate the correctness, clarity, and informativeness of both FuseFL and human explanation. Citing relevant study\cite{zhang2022itiger}, we sampled 30 faulty code files from our dataset. We found that in 22 cases, FuseFL generated the correct explanation. Furthermore, we achieved high informativeness and clarity scores of 5.7 and 5.9, respectively, on a 7-level Likert scale. These results are on par with human-generated explanations shown by no statistically significant differences between the scores.

The main contributions of our study are as follows:
\begin{itemize}
    \item We provide a dataset of faulty code, along with an explanation of why the specific line is considered faulty.
    \item We proposed the FuseFL which combines several pieces of information (spectrum-based fault localization results, test case execution outcomes, and an explanation of what the given code is intended to do) so LLM can generate better fault localization results.
    \item We conducted automatic and manual evaluations of our fault localization and generated explanations results.
   
\end{itemize}

\section{Background and Related Work}
\label{sec:background}

\subsection{Fault Localization} 
One of the most popular families in fault localization is the spectrum-based fault localization (SBFL) technique~\cite{wong2016survey}. SBFL techniques utilize a statistical formula to determine the part of the code that is responsible for a fault. It uses the test results and the execution traces to calculate the suspiciousness score, which is then arranged in descending order. 
Over the years, many SBFL techniques have been proposed~\cite{wong2013dstar,abreu2006evaluation}.
Among these, there are five most popular and well-studied SBFL techniques that were highlighted by previous study~\cite{pearson2016evaluating, widyasari2022real}, including Tarantula~\cite{jones2001visualization}, Ochiai~\cite{abreu2006evaluation}, OP2~\cite{naish2011model}, Barinel~\cite{abreu2009spectrum}, and DStar~\cite{wong2013dstar}. The formulas to calculate the suspiciousness score of each line of code $l$ of these techniques are:
\noindent\resizebox{0.5\textwidth}{!}{  \begin{tabular}{ll}
 \rule{0pt}{4ex}
$ Ochiai(l) = \frac{n_f(l)}{\sqrt{n_f.(n_p(l)+n_f(l))}} $ &  $ Tarantula(l) =  \frac{\frac{n_f(l)}{n_f}}{\frac{n_f(l)}{n_f}+\frac{n_p(l)}{n_p}} $\\
 \rule{0pt}{4ex}
 $ DStar(l) = \frac{n_f(l)^2}{n_p(l)+(n_f - n_f(l))} $ & $ Barinel(l) = 1-\frac{n_p(l)}{n_p(l) + n_f(l)}$ \\
 \rule{0pt}{4ex}
 $ OP2 = n_f(l) - \frac{n_p(l)}{n_p + 1} $ & 
\end{tabular}}
\vspace{0.1cm}

\noindent where \(n_f\) and \(n_p\) are the total number of failing and passing test cases, respectively. \(n_f(l)\) and \(n_p(l)\) are the number of failing and passing test cases executing line l, respectively.

Besides the SBFL techniques~\cite{wong2013dstar}, there have been more works proposed on fault localization~\cite{sohn2017fluccs, laghari2018use, widyasari2022xai4fl,zeng2022fault,wu2023large,kang2023preliminary}. 
Sohn et al.~\cite{sohn2017fluccs} extended SBFL by incorporating additional input factors, including code and change metrics. These factors were combined with the suspiciousness score from SBFL to create a feature for the machine learning (ML) algorithm.
Meanwhile, Laghari et al.~\cite{laghari2018use} extended SBFL by applying sequence mining to analyze a series of method calls. Zeng et al.~\cite{zeng2022fault} introduced SmartFL, a fault localization technique that leverages probabilistic methods to model program semantics. Widyasari et al.~\cite{widyasari2022xai4fl} proposed XAI4FL which utilizes the eXplainable Artificial Intelligence (XAI) technique on the ML model that classifies whether the program spectra of the element will pass or fail. Furthermore, the study by Wu et al.~\cite{wu2023large} provides prompts to instruct ChatGPT to perform fault localization. Kang et al.~\cite{kang2023preliminary} proposed AutoFL that utilizes LLM to provide fault localization results. 
In our study, we utilize the five popular SBFL techniques (i.e., Tarantula, Ochiai, DStar, Barinel, and OP2) for comparison with FuseFL, as they have been commonly used in recent studies~\cite{pearson2016evaluating, yang2021evaluating, HSFL,wu2023large,zeng2022fault,kang2023preliminary}. We also include XAI4FL~\cite{widyasari2022xai4fl}, a recently proposed SBFL tool employing XAI, in our comparison. XAI4FL relies on a decision tree for its ML model which requires the fault to have both passing and failing test cases. In instances where this condition is not met, we default to using the results obtained from the SBFL technique (Ochiai). Additionally, in Section~\ref{sec:discussion_prompt} we also compared the proposed LLM-based approach from Wu et al.~\cite{wu2023large} with our approach.

\subsection{Explanation in Fault Localization Results} 
As emphasized in the study by Kochhar et al.~\cite{kochhar2016practitioners}, fault localization needs to provide a rationale. However, there is still limited study that provides a direct explanation of the faulty line (i.e., explainable fault localization). Whyline~\cite{ko2004designing} provides an interrogative debugging approach, which enables developers to inquire about the code's runtime failure. 
Furthermore, recent studies by Wu et al.~\cite{wu2023large} and Kang et al.~\cite{kang2023preliminary} have leveraged LLM such as ChatGPT to generate explanations for fault localization. However, these studies have not included an evaluation of the generated explanations. In our study, we build a dataset of explanations, specifically for identifying the reasons behind faulty lines. 
Subsequently, we evaluated the generated explanations using both automated and manual methods. We also proposed the usage of more information on the prompts (e.g., SBFL results and explanation of the intended purpose of the code) compared with the previous studies.

\section{FuseFL}
\label{sec:approach}
\begin{figure}[htp]
\centering
\includegraphics[width=0.5\textwidth, angle=0]{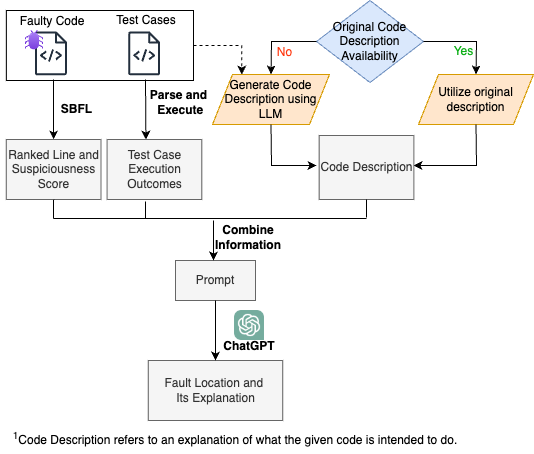}
\vspace{-0.8cm}
\caption{Proposed Approach}
\label{fig:architecture}
\end{figure}

In this section, we describe our approach (i.e., FuseFL) to generate an explanation for fault localization results. 
The overview of the proposed approach is highlighted in Figure~\ref{fig:architecture}.
Our approach leverages the strength of LLM along with multiple contextual information related to a fault. We utilize ChatGPT as the LLM model because it shows impressive performance in many fields~\cite{xia2023automated, liu2023refining, feng2023investigating, fu2023chatgpt}. We specifically utilize a version of ChatGPT-Sept-11-2023~\cite{gptversion} which uses the GPT3.5 model~\cite{brown2020language}.
Furthermore, information that we consider are suspicious line ranking obtained from SBFL results, input and output obtained from the test cases, and the code description (i.e., explanation of the intended purpose of the provided code). 
We combine these various pieces of information as inputs to the LLM to enhance the fault localization results. 

\textcolor{violet}{
\noindent\textbf{SBFL Results:} To obtain coverage information during test case execution, we utilize the Python coverage library~\cite{python_coverage}. Subsequently, we apply Spectrum-Based Fault Localization (SBFL) techniques to calculate the suspiciousness score for each line in the faulty code. These lines are then ranked in descending order based on their suspiciousness scores. In our prompts, we focus on utilizing the top 5 most suspicious lines as the primary information. In cases when the lines of code are fewer than 5, we include all available lines in our analysis.}

\textbf{Test Results:}  \textcolor{violet}{We provide several pieces of information regarding the test results, including input values, expected and actual output, and any associated errors. 
The input and expected output are extracted from the "assert" statements within the test cases. For instance, consider the following "assert" statement: \textit{"assert remove\_extras([1, 1, 3]) == [1, 3]"}. From this statement, we can conclude that the input is [1, 1, 3], and the expected output is [1, 3].
To obtain the actual output, we execute the code using the aforementioned input and then parse the resulting output. If the code cannot be executed successfully, we document the type of error encountered and specify the line where the error occurs. Meanwhile, if the code runs without issues, we record the actual output.}

\textbf{Code Description:}  
The code description outlines the requirements for the code, including its intended functionality, expected inputs, and desired outputs. An example of a code description is \textit{"The `top\_k` function takes in a list of integers as the input and returns the greatest k number of values as a list, with its elements sorted in descending order."}. In this study, we utilize the code description that is available in \textit{Refactory} dataset~\cite{hu2019re}. If FuseFL users lack code descriptions, they can utilize ChatGPT to generate one by using the following prompt: \textit{"Provide a short code description of the following code: $\langle$code$\rangle$. The provided code is expected to pass these test cases: $\langle$test cases$\rangle$"}. To validate the prompt, we run an experiment to generate the code description in our dataset and subsequently run FuseFL using these descriptions. The faults that are localized by FuseFL using the manually written and generated code description in Top-1 are 197 and 196 respectively. This minimal difference suggests that the automatically generated code description can also be used for FuseFL.

\begin{figure}[hbtp]
    \includegraphics[scale=0.53]{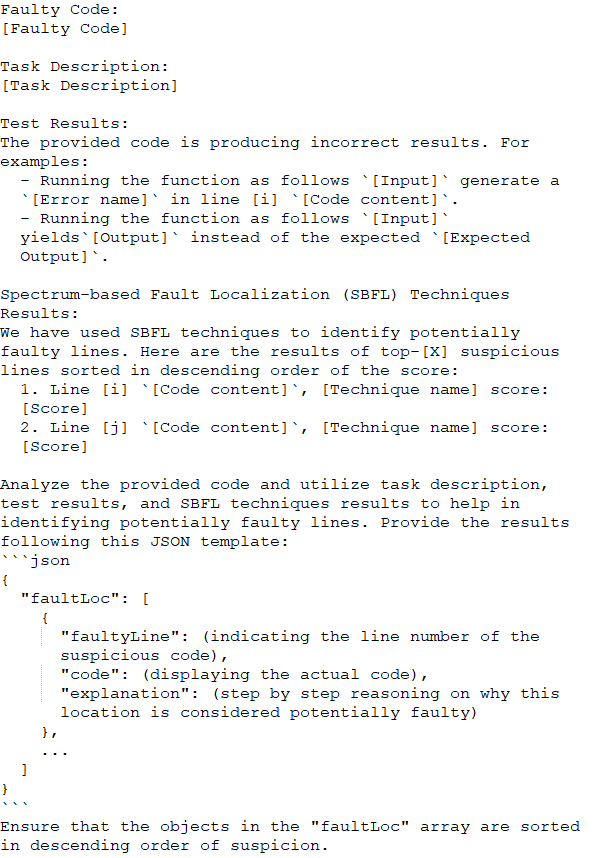}\\
    \vspace{-0.4cm}
    \caption{\textcolor{blue}{Template of generated prompt}}
    \label{fig:prompt}
    \vspace{-0.3cm}
\end{figure}

\smallskip
\textcolor{violet}{In the prompt instruction, we use a chain of thought prompting~\cite{wei2022chain,kojima2022large} and follow the approaches from previous studies that proposed prompts for ChatGPT~\cite{zhou2022large,xia2023keep,wu2023large,kang2023preliminary}. Previous studies have shown that chain of thought prompting can enhance the performance of ChatGPT~\cite{wei2022chain,kojima2022large}. 
\textcolor{blue}{Specifically, we adopted Zero-shot CoT~\cite{kojima2022large} prompting which is follow up from CoT prompting~\cite{wei2022chain}. The Zero-shot CoT~\cite{kojima2022large} is done by appending the words "Let's think step by step" to the end of the prompt. In this study, we specifically highlight that ChatGPT needs to provide "step-by-step reasoning on why this location is considered potentially faulty".}
The template prompts can be found in Figure~\ref{fig:prompt}.
To underscore the effectiveness of the prompt, we include an ablation study in Section~\ref{sec:discussion_prompt}, comparing outcomes with non-chain of thought prompting and prompt information removal.
}

\section{Experiment Setting}
\label{sec:data}

\subsection{Dataset Collection}\label{sec:dataset}
In this study, we utilized faulty codes taken from \textit{Refactory} dataset~\cite{hu2019re}. They provided 1,783 real-life faulty Python code submissions from 361 students, which were made for an introductory programming course. These submissions are valuable for fault localization as they represent the work of novice developers, often containing errors, which aligns with the practical scenarios where feedback on the code is crucial.

Furthermore, these Python faulty codes from the student assignments do not require a deep understanding of the project context. This is in contrast to other fault localization datasets such as 
Defects4J~\cite{just2014defects4j}, BugsJs~\cite{gyimesi2019bugsjs}, and BugsInPy~\cite{widyasari2020bugsinpy}, which are a collection of faulty Java, Javascript, and Python code, respectively. 
In these datasets, explaining the faultiness of a particular line of code often demands an understanding of the software projects' intricate details. 
This frequently involves the application of domain-specific knowledge and time-consuming analysis. 
The code snippets submitted by students in programming assignments are isolated, self-contained, and built specifically for novice developers, which can simplify the process of analyzing and localizing faults, requiring less extensive domain expertise. Extending the dataset with information on why specific lines of code are faulty, would improve the usability of the dataset. 
For example, the explanation of the faulty code can be utilized by students to help them understand better the reasons why certain lines are faulty.

\subsubsection{Fault Localization}
In this paper, we draw the statistically significant sample of faulty code from the \textit{Refactory} dataset~\cite{hu2019re} at a 95\% confidence level with a 5\% margin of error. We use this sampling method following the procedure used by Gunawerdana et al. in their code review analysis study~\cite{gunawardena2023concerns}.  Using the above criteria, we sampled 324 faulty Python code. For these faulty codes, we build test cases that check whether the code produces the expected output from the specific input.  
We find that the previous study~\cite{hu2019re} provided the results from automated program repairs for the faulty code. However, not every faulty code comes with a fixed version, and during our analysis, we also identified instances where the fixed version included non-essential cosmetic changes that did not directly address the underlying fault.
Because of this, we opted to manually create the fixed version of the faulty code.

When writing the fixed version, our primary objective was to introduce minimal changes to the faulty code, limited to only the necessary modifications required to fix the fault. Subsequently, we executed test cases on the fixed version to ensure it passed all of them. To identify the faulty lines, we compared the fixed version with the faulty version. During the process of fixing the code, we identified instances of faults of omission. A fault of omission occurs when certain code elements are missing, which makes the code faulty. An example of a fault of omission is a missing return statement. In cases where we encountered a fault of omission, we also specified the other potential locations to add the missing code.

\subsubsection{Explanation for the Fault Location}
Constructing an explanation dataset for a fault location is a crucial step in enhancing the interpretability of fault localization results. 
From the faulty lines and fixed version that we got from the previous steps, we explain why the specific line is faulty. The explanation of the faulty code is based on step-by-step reasoning. An example of the explanation is shown in Figure~\ref{fig:high_score_example}, the explanation reads as \textit{``This line tries to iterate over "i" using "for" loop. However, because "i" is an integer, it will lead to TypeError: "int" object is not iterable''}. In this example, we first provide information about the code's intended functionality, followed by an explanation of the correlation between the code and the potential error.
Two authors would create the explanation for the faulty lines of code. Then, another author would check the correctness of the explanation. If the explanation is deemed incorrect or insufficient, the authors will have a discussion and fix the explanation. 

\smallskip 
The total number of faulty code files in our dataset is 324. 
On average, each faulty code contains 1.85 faulty lines, totaling 600 faulty lines.  
Each faulty line is accompanied by an explanation of why the line is faulty. Despite not requiring an understanding of the project context, constructing this explanation dataset still required significant effort, taking approximately 270 hours. The explanation dataset is made publicly available as part of our replication package~\cite{replication_package}.

\subsection{Evaluation Metrics}
\subsubsection{Effectiveness of Fault Localization}
\textcolor{violet}{For evaluating the effectiveness of the fault localization technique in localizing faults, we utilize the Top-K metrics that have been used in many previous studies~\cite{parnin2011automated,kochhar2016practitioners,pearson2016evaluating,wong2013dstar, ju2014hsfal, tien, widyasari2022xai4fl}. Top-K score is the number of faulty codes for which the fault localization successfully identified the fault within the top K positions.
Given that the dataset includes multiple lines of fault, we adopted the best-case scenario~\cite{pearson2016evaluating}. In this scenario, the localization of the fault is considered successful if any of the faulty lines are correctly identified within the Top-K most suspicious lines. In cases where multiple lines share the same level of suspiciousness, we calculate the average rank~\cite{scipy} following the previous studies~\cite{wong2016survey,pearson2016evaluating}. 
Specifically, we use Top-1 as our main metric, given its critical role in automated program repair, which often assumes the perfect fault localization~\cite{chen2019sequencer,lutellier2020coconut,jiang2021cure}. In addition to the Top-1 metric we also utilize Top-2 and Top-3 as additional metrics. 
}

\subsubsection{Automated Metric for Evaluating Explanation}
\textcolor{violet}{ 
For the automated assessment of the explanations, we employ BLEURT~\cite{sellam2020bleurt}. BLEURT is an evaluation metric designed for evaluating natural language generation tasks, which is based on BERT~\cite{devlin2018bert}. In this study, we utilize the latest model which is BLEURT-20 as it is highlighted to be more accurate compared to the previous version~\cite{bleurt_github}. 
This metric has been used in previous studies~\cite{clinciu2021study,schuff2021does} that analyze the explanation. Previous study~\cite{clinciu2021study} highlights that BLEURT score has a higher correlation with human explanation compared to other metrics such as BLEU~\cite{papineni2002bleu} and ROUGE~\cite{lin2004rouge}. BLEURT provides a robust and standardized framework for quantifying the quality of generated explanations by comparing them to reference explanations. 
Considering multiple references available for the faulty line, we calculate the average BLEURT score using the highest score among the references. 
}
\textcolor{violet}{The BLEURT score typically falls within a range of 0 to 1. A score of 1 signifies that the generated explanation is of high quality, closely resembling a human-written reference. Meanwhile, a score of 0 indicates a random output from the generated text. }

\subsubsection{Manual Assessment Metric for Evaluating Explanation}\label{sec:manual_metric}
\textcolor{blue}{ 
For our manual assessment of explanations \textcolor{blue}{from humans and FuseFL}, we utilize multiple metrics: correctness (the accuracy of explaining why the line is faulty), clarity (the ease of understanding without confusion), and informativeness (the relevance and usefulness of the information provided).
To assess correctness metrics, we used binary values: "Yes" for correct explanations and "No" for incorrect ones.} 
\textcolor{violet}{If evaluators are uncertain about the correctness of a statement, they have the option to cross-reference it with online sources for verification. Furthermore, to assess clarity and informativeness, we utilized a 7-level Likert scale, where clarity ratings ranged from 1 (Very Unclear) to 7 (Very Clear), and informativeness ratings ranged from 1 (Very Not Informative) to 7 (Very Informative). These metrics have been used in previous studies~\cite{huang2023ischatgpt, clinciu2021study} that evaluated a generated explanation. 
Integrating clarity and informativeness metrics allowed us to get a more comprehensive understanding of the generated explanations' quality. 
}

\subsection{Research Questions}
\subsubsection{RQ1} \textit{How effective is FuseFL in identifying the correct location of faults? }

To assess the effectiveness of FuseFL in accurately identifying fault locations, we conduct a comparative analysis with the results obtained from five widely adopted SBFL techniques (i.e., Tarantula, Ochiai, OP2, Barinel, and DStar) and XAI4FL~\cite{widyasari2022xai4fl} which is the recent enhancement of SBFL. We have opted for these SBFL techniques because they are utilized as baselines in recent research~\cite{pearson2016evaluating, yang2021evaluating, widyasari2022xai4fl, widyasari2022real,wu2023large,zeng2022fault,kang2023preliminary}. Additionally, we have included a more recent technique (LLM-based approach) proposed by Wu et al.~\cite{wu2023large} as another baseline (see Section~\ref{sec:discussion_prompt}).

To evaluate their performances, we use Top-K score metrics.
Subsequently, we compare the results using Wilcoxon rank-sign test~\cite{wilcoxon1992individual} following previous studies~\cite{wu2023large, pearson2016evaluating}. This test is utilized to determine whether there are statistical significance differences in the effectiveness of the techniques. We apply a significance level of 1\%. This signifies that if the p-value falls below 0.01, we can reject the null hypothesis and confirm the presence of a statistically significant difference between the techniques.
Furthermore, we also compute Cohen's d effect size~\cite{becker2000effect, cohen2016power}  following -~\cite{yu2017bayesian}.  
By using Cohen's d effect size, we can check how substantially different the SBFL techniques result with FuseFL.
For interpreting the effect size, we adopt the following interpretation: \textit{negligible} if d $<$ 0.2; \textit{small} if 0.2 $\leq$ d $<$ 0.5, \textit{medium} if 0.5 $\leq$ d $<$ 0.8, and \textit{large} if d$>=$0.8~\cite{cohen2016power}.

\subsubsection{RQ2} \textit{How accurate are the explanations generated by FuseFL for the identified faulty locations?}

\textcolor{violet}{
For the second research question, our goal was to compare the explanations made by humans and generated by FuseFL. This evaluation is instrumental in determining how well FuseFL can explain the faulty code and whether its explanations align with those crafted by humans. 
Accurate explanations are crucial for developers, as they aid in comprehending code issues and facilitate effective debugging.  
In this RQ, we utilize the automated metric BLEURT to evaluate the generated explanation. The score from BLEURT measures the quality and fluency of generated text in comparison to reference text (i.e., human explanation). 
}

\subsubsection{RQ3} \textit{How do developers perceive FuseFL's generated explanations compared with human explanations?}

To ensure the quality of the explanations aligns with developer expectations, we conducted a manual evaluation involving a diverse group of 10 evaluators with varying years of programming experience. On average, these evaluators possessed 7 years of programming experience, with a minimum of 4 years and a maximum of 10 years. All evaluators were familiar with fault localization and code using Python. The participant recruitment and evaluation procedures were conducted following the Institutional Review Board (IRB) approval, identified by reference number IRB-23-163-A114(1023). To facilitate the evaluation process, we utilized an online form (i.e., Google Form), allowing evaluators to complete it at their convenience.

In this evaluation, we randomly sampled 30 faulty code samples from the dataset, in line with the approach used in a previous study~\cite{zhang2022itiger} to assess the issue titles through manual evaluation. 
We presented the evaluators with the faulty code (faulty line highlighted), along with both the automated generated and human explanations. To mitigate bias, we did not disclose how the explanations were generated. Additionally, we randomized the order of explanations to further reduce bias.
We instructed evaluators to assess the correctness, clarity, and informativeness of the explanation.   
For the correctness metric, we aggregate the results using a consensus strategy. We considered an explanation correct if at least 90\% of evaluators agreed. 
Meanwhile, for clarity and informativeness metrics, we averaged the scores given by all 10 evaluators.

\section{Results}
\label{sec:results}

\subsection{RQ1: How effective is FuseFL in identifying the correct location of faults?} 
The Top-K scores for SBFL techniques, XAI4FL, and FuseFL are presented in Table~\ref{fig:RQ1_topk}.
The number of faults localized at the Top-1 position by FuseFL is significantly higher compared to SBFL techniques and XAI4L. SBFL techniques localized 89 to 117 faults at the Top-1 position, while FuseFL can localize 197 faults, providing 77 to 95 more faults localized at the Top-1 position. 
FuseFL achieved a minimum improvement of 68.4\% in the number of localized faults at the Top-1 position compared to SBFL techniques (i.e., Ochiai). Including XAI4FL in the comparison, we observe that XAI4FL localized 149 faults at the Top-1 position with 27\% improvement compared to Ochiai. However, it falls behind FuseFL, which outperforms it by 32.3\%. 
Furthermore, FuseFL also showed improvement in the Top-2 and Top-3 results compared to baselines which ranged from 11\% to 37\% and  4\% to 20\%, respectively. 
As many automated program repairs assume a perfect localization (i.e., a fault that is localized at the Top-1 position)~\cite{chen2019sequencer,lutellier2020coconut,jiang2021cure}, we believe that improvement at the Top-1 score would be beneficial for practical usage.

\begin{table}[]
\centering
\caption{Top-K result for SBFL techniques and FuseFL}
\label{fig:RQ1_topk}
\begin{tabular}{c|rrrrr|r|r|}
\cline{2-8}
\multicolumn{1}{l|}{\multirow{2}{*}{}} & \multicolumn{5}{c|}{\textbf{SBFL}}                                                                                                                                                              & \multicolumn{1}{l|}{\multirow{2}{*}{\textbf{XAI4FL}}} & \multicolumn{1}{l|}{\multirow{2}{*}{\textbf{FuseFL}}} \\ \cline{2-6}
\multicolumn{1}{l|}{}                  & \multicolumn{1}{l|}{\textbf{Tar}} & \multicolumn{1}{l|}{\textbf{Och}} & \multicolumn{1}{l|}{\textbf{OP2}} & \multicolumn{1}{l|}{\textbf{Bar}} & \multicolumn{1}{l|}{\textbf{DStar}} & \multicolumn{1}{l|}{}  & \multicolumn{1}{l|}{}                                         \\ \hline
\multicolumn{1}{|c|}{\textbf{Top-1}}   & \multicolumn{1}{r|}{89}                 & \multicolumn{1}{r|}{117}             & \multicolumn{1}{r|}{115}          & \multicolumn{1}{r|}{89}               & 115    & 149                             & \textbf{197}                                                  \\ \hline
\multicolumn{1}{|c|}{\textbf{Top-2}}   & \multicolumn{1}{r|}{175}                & \multicolumn{1}{r|}{208}             & \multicolumn{1}{r|}{208}          & \multicolumn{1}{r|}{175}              & 208         & 216                        & \textbf{240}                                                  \\ \hline
\multicolumn{1}{|c|}{\textbf{Top-3}}   & \multicolumn{1}{r|}{216}                & \multicolumn{1}{r|}{247}             & \multicolumn{1}{r|}{247}          & \multicolumn{1}{r|}{217}              & 247   & 248                              & \textbf{259}                                                  \\ \hline
\multicolumn{8}{l}{\footnotesize Tar = Tarantula; Och = Ochiai; Bar = Barinel.}
\end{tabular}
\vspace{-0.2cm}
\end{table}

\begin{table}[h]
\vspace{-0.2cm}
\caption{Cohen’s d effect sized and Wilcoxon test results}
\label{tab:RQ1_test}
\centering
\begin{threeparttable}
\centering
\begin{tabular}{|l|ccc|}
\hline
\multicolumn{1}{|c|}{}                                     & \multicolumn{3}{c|}{\textbf{FuseFL}}                                                                            \\ \cline{2-4} 
\multicolumn{1}{|c|}{\multirow{-2}{*}{\textbf{Technique}}} & \multicolumn{1}{c|}{\textbf{Top-1}}                    & \multicolumn{1}{c|}{\textbf{Top-2}} & \textbf{Top-3}  \\ \hline
Tarantula                                         & \multicolumn{1}{c|}{0.71* (M)} & \multicolumn{1}{c|}{0.42* (S)}       & 0.30* (S)       \\ \hline
Ochiai                                         & \multicolumn{1}{c|}{0.51* (M)} & \multicolumn{1}{c|}{0.21* (S)}       & 0.09 (N)       \\ \hline
OP2                                        & \multicolumn{1}{c|}{0.52* (M)} & \multicolumn{1}{c|}{0.21* (S)}       & 0.09 (N)       \\ \hline
Barinel                                         & \multicolumn{1}{c|}{0.71* (M)} & \multicolumn{1}{c|}{0.42* (S)}       & 0.30* (S)       \\ \hline
DStar                                         & \multicolumn{1}{c|}{0.52* (M)} & \multicolumn{1}{c|}{0.21* (S)}       & 0.09 (N)       \\ \hline
XAI4FL                                         & \multicolumn{1}{c|}{0.3* (S)} & \multicolumn{1}{c|}{0.16* (N)}       & 0.08 (N)       \\ \hline
\end{tabular}
\end{threeparttable}

\smallskip
\justifying
\footnotesize
\noindent
"*" indicates that the difference is statistically significant at 1\%.

\noindent "(Neglible(N)/Small(S)/Medium(M)" denotes the  effect size. 
\vspace{-0.2cm}
\end{table}

\textcolor{violet}{
To further check whether the differences are statistically significant, we conducted a statistical test comparing the results between SBFL techniques and FuseFL. The results are shown in Table~\ref{tab:RQ1_test}. 
Observing the Top-1 and Top-2 statistical test results, we found that the FuseFL outperforms SBFL techniques and XAI4FL with statistically significant differences at a 1\% significance level.
The effect size also indicates that the Top-1 score differences between all the SBFL techniques and FuseFL are substantial with a medium effect size.
At the Top-3 position, FuseFL results are comparable with the results from Ochiai, OP2, DStar, and XAI4FL. The p-value for Ochiai, OP2, and DStar is 0.052 which is only statistically significant at a 10\% significance level.}

To further analyze the results, we choose Ochiai to represent the SBFL techniques as it localized more faults at the Top-1 position. 
Among the 117 faults that are successfully localized in Top-1 by Ochiai, 116 faults (99\%) are also localized in Top-1 by FuseFL. 
This shows that the majority of faults localized by Ochiai are also successfully identified by FuseFL. 
The single fault that is uniquely localized at Top-1 by Ochiai is localized at Top-2 by FuseFL, indicating that in instances where Ochiai achieves perfect localization, FuseFL still yields strong results. 
Meanwhile, for the 81 faults that are uniquely localized at Top-1 by FuseFL, 32\% of them are localized outside of Top-2 by Ochiai. 
We also found that the SBFL techniques results have many lines that share the same suspiciousness score. If the scores are shared among several lines, the final position would be averaged, following the common practice from previous studies~\cite{pearson2016evaluating,wong2016survey}. For example, if the highest suspiciousness score is shared between two lines, the final position would be 1.5. 
This would make the perfect fault localization not achievable for SBFL techniques in such cases.

\begin{tcolorbox}
{\textbf{RQ1 Findings:} 
FuseFL improves the number of faults localized in Top-1, Top-2, and Top-3. 
Furthermore, in terms of Top-1, FuseFL statistically significantly outperforms the best-performing baseline (XAI4FL) with at least a 32.3\% improvement. 
}
\end{tcolorbox}

\subsection{RQ2: How accurate are the explanations generated by FuseFL for the identified faulty locations?}
In this RQ, we compare FuseFL-generated explanations to human explanations by applying the BLEURT metric on successfully localized faults in Top-K. 
The average BLEURT score for Top-1, Top-2, and Top-3 are 0.492, 0.486, and 0.486 respectively.  
We found that the highest BLEURT score is achieved in the Top-1 results, where the score is 1.2\% higher compared to the score achieved in the Top-2 and Top-3 results.

\begin{figure}[h]
\includegraphics[scale=0.596]{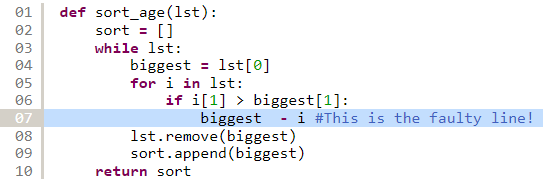}\\
\noindent\fbox{%
    \parbox{0.48\textwidth}{%
        Human Explanation:\\
This line is executed when the condition of "i[1] > biggest[1]" is satisfied. However, the syntax of this line is incomplete, in this line, it only has "biggest  - i" without the assignment variable itself.
    }%
}

\noindent\fbox{%
    \parbox{0.48\textwidth}{%
        FuseFL Explanation:\\
The code mistakenly uses the subtraction operator instead of the assignment operator. Change "biggest  - i" to "biggest = i" to correctly update the value of "biggest" with the current tuple "i" if it has a higher age.
    }%
}
\caption{Example of the BLEURT low score explanation}
\label{fig:low_score_example}
\end{figure}

We then analyze the explanation generated by FuseFL which has a lower score. Our observation found that while the explanation created by FuseFL is correct, it is expressed differently from the human explanation. An example can be found in Figure~\ref{fig:low_score_example} with line number 7 as the faulty line.  This function "sort\_age" is expected to sort the people, represented using a tuple (<gender>, <age>), based on the age and return a list of people in descending order. 
In this example, we observe that the FuseFL explanation is correct. However, the BLEURT score is low (i.e., 0.32). These low results may occur as the human explanation is focused on why the syntax is incorrect, while the FuseFL explanation is focused on why the subtraction operation is a mistake. Despite both being correct, they are focused on different aspects, resulting in a low BLEURT score. Through further analysis, we found that FuseFL tends to create explanations by providing an approach on how to fix the code, rather than reasoning on the root cause of the fault. 
In the previous example, we can see that the FuseFL highlights that the subtraction is a mistake, as it should be changed into an assignment operator (i.e., the fix needs to be applied in the code).

\begin{figure}[h]
\includegraphics[scale=0.596]{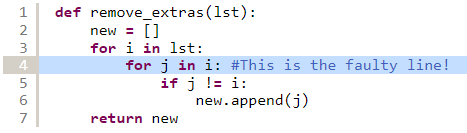}\\
\noindent\fbox{%
    \parbox{0.48\textwidth}{%
        Human Explanation:\\
This line tries to iterate over "i". However, because "i" is an integer, it will lead to TypeError: "int" object is not iterable
    }%
}

\noindent\fbox{%
    \parbox{0.48\textwidth}{%
        FuseFL Explanation:\\
The "TypeError: `int' object is not iterable" occurs because the variable "i" is an integer and cannot be iterated over.
    }%
}
\caption{Example of the BLEURT high score explanation}
\label{fig:high_score_example}
\vspace{-0.5cm}
\end{figure}

\textcolor{violet}{Furthermore, the example of FuseFL's explanation that gets a high score can be seen in Figure~\ref{fig:high_score_example}. This "remove\_extras" function takes an input list and returns a new list with all repeated occurrences of any element removed while maintaining the same order as the input list. 
In this example, FuseFL's explanation achieved a score of 0.73, as the explanation is similar to the human explanation. The explanation differences are mainly in their structure, where FuseFL's explanation discussed the error first, while the human explanation discussed the context of the code first. the human explanation.  
}

\begin{tcolorbox}
{\textbf{RQ2 Findings:} 
\textcolor{violet}{
The average of the BLEURT score in the Top-1 result is 0.492. \textcolor{blue}{We find that the seemingly low BLEURT score does not necessarily mean that the explanation is incorrect; it indicates a variance between the FuseFL-generated and human explanations.} 
}
}
\end{tcolorbox}

\subsection{RQ3:How do developers perceive FuseFL's generated explanations compared with human explanations?}
\begin{figure}[h]
\includegraphics[scale=0.596]{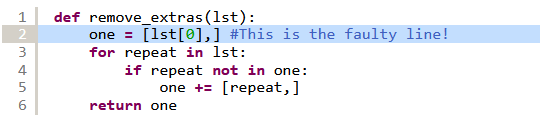}\\
\noindent\fbox{%
    \parbox{0.48\textwidth}{%
        Human Explanation:\\
The initial value of "one" is set to contain the first element of the "lst". However, if "lst" is empty, this line will throw an IndexError.
    }%
}
\noindent\fbox{%
    \parbox{0.48\textwidth}{%
        FuseFL Explanation:\\
The code attempts to initialize a new list "one" with the first element of the input list "lst". However, if the input list is empty, an IndexError occurs because there is no element at index 0.
    }%
}
\caption{Example of correct FuseFL explanation}
\label{fig:correct_case}
\vspace{-0.2cm}
\end{figure}

\begin{figure}[h]
\includegraphics[scale=0.596]{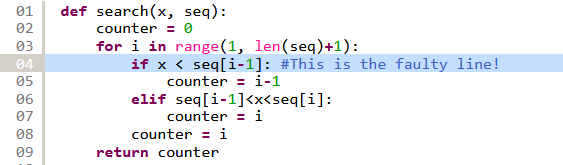}\\
\noindent\fbox{%
    \parbox{0.48\textwidth}{%
        Human Explanation:\\
The condition checks if "x" is less than previous element, if it is, "i-1"  would be assigned to "counter". However, on the line 8, the variable "counter" is assigned with "i" which overwrite the counter that is updated in the next line.
    }%
}
\noindent\fbox{%
    \parbox{0.48\textwidth}{%
        FuseFL Explanation:\\
This line incorrectly updates the counter when x is less than the current element in the sequence. It should update the counter only when x is greater than or equal to the current element.
    }%
}
\caption{Example of wrong FuseFL explanation}
\label{fig:wrong_case}
\vspace{-0.3cm}
\end{figure}

There are three metrics that we utilize for human evaluation: correctness, informativeness, and clarity. In terms of the correctness metrics, the vote results from the human evaluation show FuseFL correctly generated 22 explanations out of the 30 considered cases. These results highlight FuseFL's capability to generate correct explanations. An example of a correct explanation generated by FuseFL can be seen in Figure~\ref{fig:correct_case}. In this example, FuseFL correctly explains why the line is faulty and also highlights the input that causes the error in the code. Specifically, it highlights that the empty input list would make the code raise an IndexError as this corresponding line tries to access the element at index 0. Meanwhile, an example of a wrong explanation can be seen in Figure~\ref{fig:wrong_case}. In this example, FuseFL provided incorrect information, as it states "the current element", which is supposed to be "the previous element" as the code indicates "seq[i-1]" with "i" being the current index. Based on these findings, even though it mostly can generate the correct explanation, it is recommended that developers continue to exercise caution when leveraging the explanations generated by LLMs in general.

For the informativeness and clarity metrics, FuseFL achieves average results of 5.7 and 5.9, respectively, on a 7-level Likert scale. These scores are comparable with the scores for human explanation, which are 5.8 for informativeness and 5.9 for clarity, showing that there is only a small difference of 0.1 in the informativeness metric. The Wilcoxon rank-sign test results confirm that the disparities between FuseFL and human explanations in both informativeness and clarity metrics are statistically not significant with a p-value > 0.01.

\begin{figure}[h]
\includegraphics[scale=0.596]{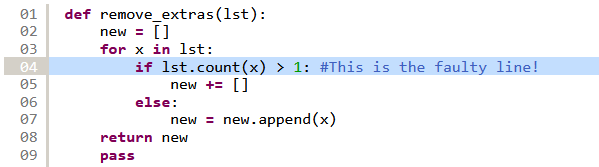}\\
\noindent\fbox{%
    \parbox{0.48\textwidth}{%
        FuseFL Explanation:\\
The line checks if the count of "x" in "lst" is greater than 1. This condition determines if an element is repeated in the list.
    }%
}
\caption{Example of FuseFL explanation with low informativeness score}
\label{fig:low_score_informative}
\end{figure}

An example case where the FuseFL explanation receives a lower score in informativeness is highlighted in Figure~\ref{fig:low_score_informative}. For this example, 
FuseFL does not give much information on why the line is deemed to be faulty. FuseFL can also generate an explanation with a higher score compared to human explanation, as can be seen in the example shown in Figure~\ref{fig:correct_case} where FuseFL achieves 0.2 higher average informativeness and clarity. The higher scores in this example can be attributed to the FuseFL providing a clearer reason for why the error occurred, which is due to the code trying to access index 0 of an empty list. Furthermore, the FuseFL score for informativeness and clarity only has a small difference of 0.2, which indicates that FuseFL can provide an effective balance of clear and informative explanations.

\begin{tcolorbox}
{\textbf{RQ3 Findings:} 
In 22 out of 30 cases, FuseFL generated correct explanations according to human evaluation. 
Furthermore, FuseFL achieves high informativeness (5.7) and clarity (5.9) scores, statistically not significantly different from human explanations.
}
\end{tcolorbox}

\section{Discussion}
\label{sec:discussion}
\subsection{Error Type and its Impact on FuseFL Results}
\textcolor{violet}{We further analyzed the FuseFL results and categorized them based on the error in the code. We want to check whether the different types of errors can affect the results of FuseFL. For this analysis, we first categorize the faulty code by applying open card sorting~\cite{spencer2009card} to categorize the error in the code. 
}
\textcolor{violet}{First, we created cards containing faulty code, test results, code descriptions, and explanations of the faulty code. 
Then, we examined each card, categorizing them into their respective groups. Once all the cards had been organized, the authors assigned names to these groups that encapsulated the patterns found in the cards. This open card sorting process aligns with practices in prior researches~\cite{xu2020reinventing, bacchelli2013expectations}. }

\subsubsection{Fault Localization Results}\label{sec:fault_res}
\textcolor{violet}{
As we want to analyze the pattern of the code where FuseFL failed on localizing, we filter FuseFL results, retrieving only faulty codes that cannot be localized within the Top-3 by FuseFL. We retrieved a total of 64 fault codes, which we further divided into two main categories: Runtime Error and Output Error. A Runtime Error is a condition when the code cannot be executed until completion (i.e., an error during code execution that causes an early exit). Meanwhile, for the Output Error, the code can be executed until completion, however, the output of the code differs from the output expected by the developer.  
There are several subcategories for both Runtime Errors and Output Errors, such as IndexError, TypeError, RecursionError, and others within the Runtime Error category. In contrast, Output Errors can be divided into issues like Missing Return Statements, Incorrect Assignments, Flawed Condition Checks, and more. However, we only discussed these main categories in the paper as the effects were more prominent.}

\begin{figure}
\centering
\includegraphics[scale=0.64]{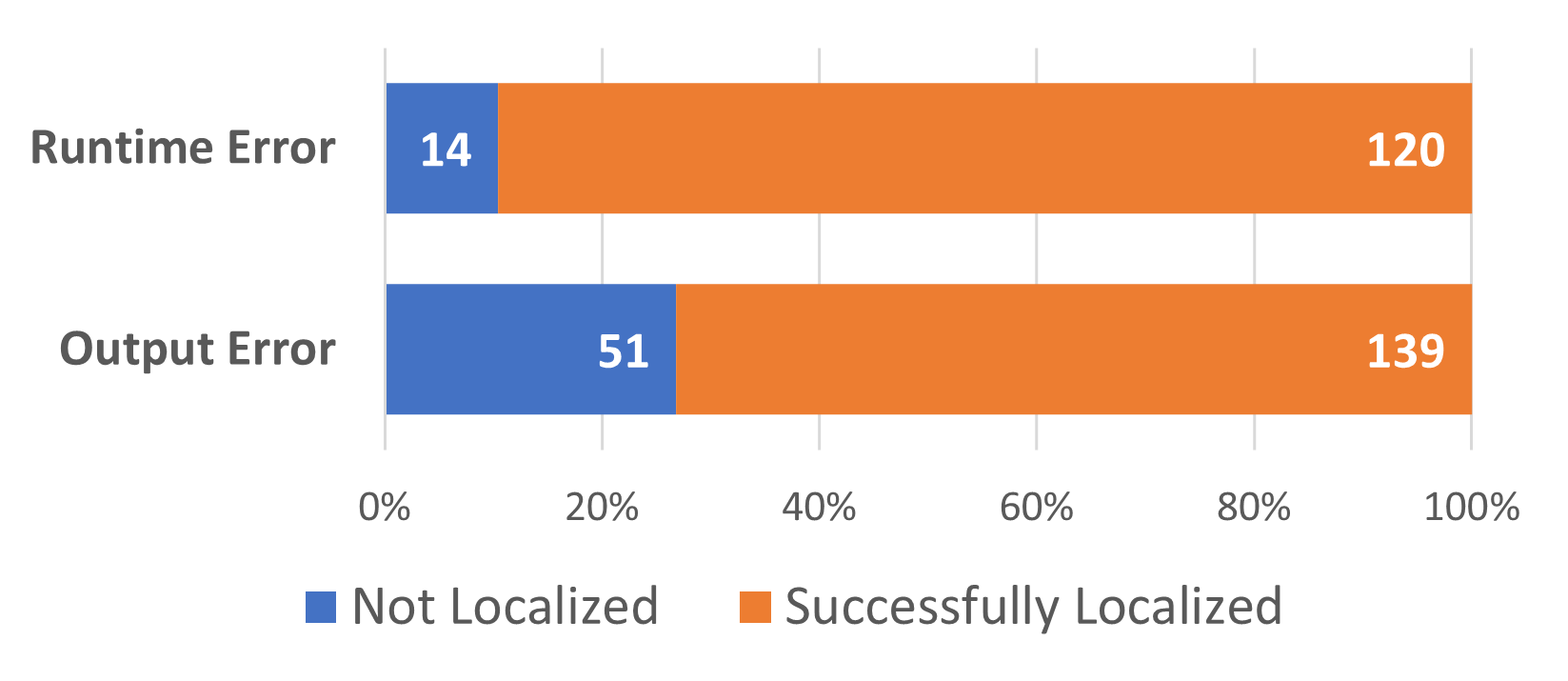}
\vspace{-0.4cm}
\caption{Error type distribution with the fault localization results.}
\label{fig:error_type}
\vspace{-0.2cm}
\end{figure}

The distribution of these two categories is shown in Figure~\ref{fig:error_type}. In cases where FuseFL failed to localize faults, a larger proportion (78.5\%) have Output Error, while the remaining 21.5\% have Runtime Error. Similarly, when considering successfully localized faults by FuseFL, the percentage of cases with Runtime Error is higher than those with Output Error. 
This shows that FuseFL does a better job when the Runtime Error is found in the code, which is typically more explicit about the error. Meanwhile, FuseFL still has the limitation of localizing the faulty code with Output Error.

\subsubsection{Explanation of Fault Location Results}
\textcolor{violet}{
We want to analyze whether there is a correlation between specific types of error with the BLEURT scores. We utilize the category that we get from analysis in Section~\ref{sec:fault_res} (i.e., Output Error and Runtime Error).
Then, we divided the faulty code explanation into two groups based on their BLEURT scores: the high-score explanation (indicating better alignment with human explanation) and the low-score explanation (indicating less correspondence with human explanation). We divide the score using its median (0.4702), where the high-score has an equal or larger score than the median. There are 98 explanations in the low-score group, while 99 in the high-score group.
}

\begin{figure}
\centering
\includegraphics[scale=0.64]{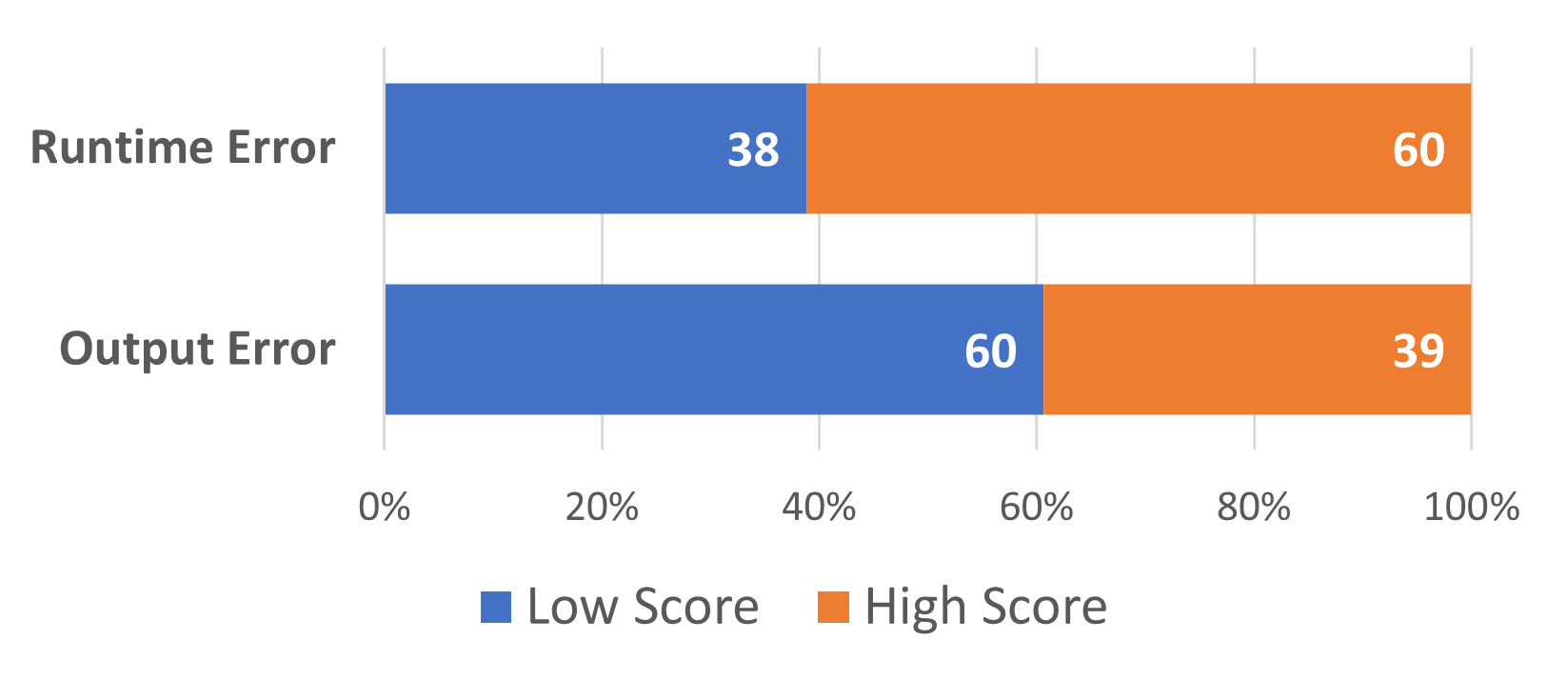}
\vspace{-0.4cm}
\caption{Error type distribution with the explanation results.}
\label{fig:error_type_2}
\vspace{-0.2cm}
\end{figure}

The distribution of the results is shown in Figure~\ref{fig:error_type_2}. Observing the distribution, we found that in the high score group, the faulty code mostly (60.6\%) has Runtime Error. 
We also found that cases containing Runtime Error typically have explanations with high BLEURT scores. This is in contrast with the explanations for faulty codes containing Output Error, where 60.6\% of the explanations have a low BLEURT score (as opposed to Runtime Error with 38.8\% low score explanations). These results are in line with the fault localization results discussed in Section~\ref{sec:fault_res}, wherein the number of faulty codes that FuseFL failed to localize is notably higher in cases involving Output Error. This highlights the importance of future research to enhance fault localization and explanation results for codes containing Output Error. One of the potential solutions is to incorporate multi-round interaction to further guide the LLMs toward a more accurate answer. 

\begin{table}[t]
\centering
\caption{Evaluation results divided by the developers experience.}
\label{fig:discussion_experience}
\begin{tabular}{|l|r|r|r|}
\hline
\multicolumn{1}{|c|}{\textbf{Developers}} & \multicolumn{1}{c|}{\textbf{Correctness}} & \multicolumn{1}{c|}{\textbf{Informativeness}} & \multicolumn{1}{c|}{\textbf{Clarity}} \\ \hline
Experienced                 & 23             & 5.6                                             & 5.5                                                  \\ \hline
Novice                      & 27        & 6.2                                              & 6.0                                                  \\ \hline

\end{tabular}
\vspace{-0.4cm}
\end{table}

\subsection{Developers' Experience and FuseFL Evaluation}
\textcolor{violet}{
Given that the explanation can be perceived differently by different developers, we also analyze the evaluation result by taking into account the developers' experience. We analyzed how developers with varying levels of expertise perceive and assess the explanations provided. Understanding these differences is essential for tailoring the explanation for fault localization to meet the specific needs and expectations of both experienced and novice developers.}

\textcolor{violet}{We divide the developers into novice and experienced based on their years of experience. Developers with years of experience equal to or more than the median number (i.e., 8 years) would be considered experienced, while the rest would be considered as novice. The number of FuseFL-generated explanations that are assessed as correct is higher for novice developers compared to experienced developers. 
Furthermore, for the clarity and informativeness score of FuseFL explanations, the experienced developers provide lower scores compared to the novice developers. The results can be found in Table~\ref{fig:discussion_experience}. During the evaluation, the developers are also given open-ended questions regarding their thoughts on the explanations that are being provided. Several experienced developers mentioned that some of the explanations are too long and they would prefer shorter explanations. They also shared that such long explanations might be more helpful for novice developers. These reasonings provide some insights into why experienced developers give lower scores for clarity and informativeness compared to novice developers. In future studies, we can survey more developers regarding their preferences on the explanation and provide an explanation that can be tailored based on their preferences.}

\begin{table*}[]
\centering
\caption{The Top-K results by using different prompts on ChatGPT}
\label{tab:prompt_exp}
\begin{tabular}{|l|r|r|r|r|r|r|r|r|r|}
\hline
               & \multicolumn{1}{c|}{\textbf{Baseline}~\cite{wu2023large}} & \multicolumn{1}{c|}{\textbf{CoT}} & \multicolumn{1}{c|}{\textbf{TestRes}} & \multicolumn{1}{c|}{\textbf{SusScore}} & \multicolumn{1}{c|}{\textbf{CodeDesc}} & \multicolumn{1}{c|}{\textbf{SusScore+TestRes}} & \multicolumn{1}{c|}{\textbf{TestRes+CodeDesc}} & \multicolumn{1}{c|}{\textbf{SusScore+CodeDesc}} & \multicolumn{1}{c|}{\textbf{FuseFL}} \\ \hline
\textbf{Top-1} & 150                                    & 163                               & 187                                   & 169                                    & 180                                        & 192                                            & 190                                                & 177                                             & \textbf{197}                                \\ \hline
\textbf{Top-2} & 190                                    & 197                               & 232                                   & 217                                    & 211                                        & 231                                            & 231                                                & 222                                             & \textbf{240}                                \\ \hline
\textbf{Top-3} & 216                                    & 214                               & 242                                   & 257                                    & 230                                        & 256                                            & 246                                                & 255                                             & \textbf{259}                                \\ \hline
\end{tabular}
\vspace{-0.2cm}
\end{table*}

\subsection{ChatGPT Prompts Experiments}\label{sec:discussion_prompt}
To get the prompts that we used in the study highlighted in Section~\ref{sec:approach}, we conducted experiments with several prompt variations. The results of these prompt experiments are presented in Table~\ref{tab:prompt_exp}. Initially, we adopted a baseline prompt from the LLM-based approach used in Wu et al. study~\cite{wu2023large}, we called this a "Baseline" prompt. 
Then, we enhance the prompts and utilize Chain of Thought (CoT) prompting by specifically instructing ChatGPT to do a step-by-step reasoning, we referred to this prompt as a "CoT" in Table~\ref{tab:prompt_exp}.
From the comparison between the results of these two approaches, we find that CoT prompting provided a higher number of successfully localized faults at Top-1 and Top-2. Moreover, when assessing the generated explanations, CoT prompting (0.490) received a higher BLEURT score compared to the baseline (0.471). 
Given the higher performance in both fault localization and explanation score, we further enhanced this CoT prompt by incorporating additional information.

Combining the additional information with the CoT prompt, in total, we have 7 different prompts: "TestRes" (include the test results), "SusScore" (include the SBFL results), "CodeDesc" (include the code description), "SusScore+TestRes" (include the SBFL and test results), "TestRes+CodeDesc" (include the test results and code description), "SusScore+CodeDesc" (include the SBFL results and code description), and "FuseFL" (include all the information). 

Comparing the outcomes of the SusScore, TestRes, and CodeDesc prompts, we observed that the TestRes prompt achieved the highest number of localized faults at Top-1 and Top-2. Meanwhile, the highest number of faults that are localized at Top-3 come from the SusScore prompt. Notably, the SusScore prompt itself exhibited a 44\% improvement over SBFL results at Top-1. However, it had a comparatively minor impact on the base CoT prompting, contributing only a 3.7\% increase at Top-1. This observation implies a potential correlation with the limited success of SBFL techniques in localizing faults at Top-1 and its better performance at lower positions (Top-3). Every fault that is successfully localized at Top-1 by SBFL is also localized at Top-1 using the "SusScore" prompt. In 77 cases where SBFL fails to localize a fault at the Top-3 position, only 8 faults are localized at Top-1 using the "SusScore" prompt. 
Utilizing only SBFL information, without the incorporation of additional data, might not yield substantial benefits for perfect fault localization.

In our comparison of prompts containing more information, namely "SusScore+TestRes", "TestRes+CodeDesc", and "SusScore+CodeDesc", we observed that "SusScore+TestRes" outperformed the other two prompts in terms of the total number of successfully localized faults at Top-1, Top-2, Top-3. This result aligns with the result from single information, where TestRes yielded the highest scores for Top-1 and Top-2, while SusScore provided the highest scores for Top-3.
Utilizing all available information in the prompt, as in FuseFL we get the highest scores across all the considered Top-K values compared to other prompts. Therefore, we opted to use all available information in our prompt. Furthermore, comparing FuseFL with the "Baseline"~\cite{wu2023large} we get an improvement of 33\% fault localized at Top-1. 
These results highlight that a combination of information can enhance LLM results.

\subsection{LLM Token Limit on FuseFL Evaluation}\label{sec:discussion_token_limit} 
In this section, we discuss the effect of the LLM token limit in our evaluation. 
The default token limitation for freely accessible GPT-3.5 was originally 4,096 tokens. Meanwhile, the paid-to-use models such as GPT-4 allow up to 32,768 tokens~\cite{gptdocumentation}. 
In our evaluation, the average number of tokens shared between the prompt and completion is 1,293, which falls below the token limitation.
This indicates that ChatGPT can handle our data well. This stems from the fact that the data is submitted by students in programming assignments, which are simple, isolated, and self-contained. In cases where the codebase is more complex, further data processing strategy would be required to provide sufficient context.

\section{Threats to Validity}
\label{sec:threats}
An internal validity threat arises from the explanation of the faulty line that we created. Despite our best efforts, there remains a chance that we might generate inaccurate explanations. 
To address this concern, we have included our explanation dataset in the replication package. This allows other researchers to validate our findings and assess the accuracy of our explanations.

In our human evaluation, we use the binary correctness metric, to distinguish correct and incorrect explanations, and utilize a consensus strategy to aggregate the results. While this method may result in occasional evaluator disagreements, we mitigate this potential issue by setting a high cutoff (90\%) to ensure that the disagreement rate remains minimal.

Another threat comes from the generalizability of our findings.  Given our specific focus on faulty code originating from Python assignments submitted by students, there may be limitations in generalizing our results to other datasets with greater complexity (e.g., those requiring knowledge of project-specific information). \textcolor{blue}{We selected \textit{Refactory} dataset~\cite{hu2019re} because it allows for better control over the quality of generated explanations, as it does not require specialized domain expertise to generate accurate explanations. In contrast, generating well-crafted and accurate explanations for complex code necessitates domain expertise.} 
In future research, we intend to expand our analysis to encompass a wider range of faulty code with varying levels of complexity and difficulty.

Due to the rapid evolution of ChatGPT, the output generated by the current prompt may exhibit variations across different versions. This study utilized the most recent freely available version of ChatGPT (ChatGPT-Sept-11-2023~\cite{gptversion}) that is based on the GPT3.5 model~\cite{brown2020language}. We opted for GPT-3.5 due to its greater accessibility compared to GPT-4. As we take the faulty codes from previous study~\cite{hu2019re}, ChatGPT may have seen it. However, we manually label the fault location and build the explanation of the faulty line from scratch. Thus, ChatGPT would not have seen these in their training. 

\section{Conclusion and Future Direction}
\label{sec:conclusion}
We conduct an investigation on the effectiveness of explainable fault localization. We proposed FuseFL which utilizes LLM, along with additional fault-related information (i.e., SBFL results, test results, and code description) for fault localization and to generate step-by-step reasoning about the fault. 
Our evaluation results show that FuseFL statistically significantly outperforms baselines in localizing faults at Top-1. 
Furthermore, to evaluate the explanation generated by FuseFL, we created a dataset of human explanations that provide step-by-step reasoning on why specific lines of code are considered faulty. Comparing FuseFL's generated explanations to human explanations, it achieved an average BLEURT score of 0.492 for successfully localized Top-1 faults. We also conducted a human evaluation where we found that FuseFL provides 22 correct explanations out of 30 randomly sampled cases. 
Based on these results, FuseFL shows that combining several pieces of information can help to improve the performance of LLM in explainable fault localization.

For future work, we plan to further improve the quality of the explanation for fault localization. We will explore multi-round interactions with LLM in order to support richer dialogues. We also plan to investigate the performance of FuseFL in a more complex codebase. 

\section*{Acknowledgement}
This research / project is supported by the National Research Foundation, under its Investigatorship Grant (NRF-NRFI08-2022-0002). Any opinions, findings and conclusions or recommendations expressed in this material are those of the author(s) and do not reflect the views of National Research Foundation, Singapore.

\balance
\bibliographystyle{IEEEtran}
\bibliography{main}

\end{document}